\documentclass[12pt,a4paper,twoside,fleqn]{article} 
\usepackage{latexsym} 
\usepackage{emlines} 
\usepackage{epsf} 
\usepackage{feynmf} 
 
\begin{document} 

\newcommand{\sumint}[1]
{\begin{array}{c}
\\
{{\textstyle\sum}\hspace{-0.9em}{\displaystyle\int}}\\
{\scriptstyle{#1}}
\end{array}} 

\newcommand{\ket}[1]{\left|#1\right>} 
\newcommand{\bra}[1]{\left<#1\right|} 
\newcommand{\dotA}{\stackrel{\;\;.}{\vec A}}

\begin{center} 
{\Large \bf Quantum radiation in external background fields} 
\end{center} 
\normalsize 
\bigskip 
\begin{center} 
 
\vspace{0.3cm} 
{\large Ralf Sch\"utzhold, G\"unter Plunien and Gerhard Soff} 
  
{\sl Institut f\"ur Theoretische Physik, Technische Universit\"at Dresden,\\ 
Mommsenstr. 13, D-01062 Dresden, Federal Republic of Germany}  
 
\end{center}

\vspace{1.0cm} 
 
\begin{abstract} 
A canonical formalism is presented which allows for investigations of  
quantum radiation induced by localized, smooth disturbances of
classical background fields by means of a perturbation theory approach.
For massless, non-selfinteracting quantum fields at zero temperature we
demonstrate that the low-energy part of the spectrum of created particles 
exhibits a non-thermal character. Applied to QED in varying dielectrics the 
response theory approach facilitates to study two distinct processes
contributing to the production of photons: the squeezing effect due to 
space-time varying properties of the medium and of the velocity effect
due to its motion. The generalization of this approach to finite temperatures
as well as the relation to sonoluminescence is indicated.
 
\end{abstract} 
 
\bigskip 
 
PACS-numbers:  12.20.; 42.50.; 04.60.Ds; 78.60.Mq\\  
\vspace{0.1cm}


\section{Introduction} 

Perhaps one of the most fascinating aspects of the static Casimir effect
\cite{s1,lam97}
is that it demonstrates the highly non-trivial nature of the vacuum state
in quantum field theory (see Refs. \cite{s2,s3,s4} for reviews). 
During the last decade the dynamical Casimir effect, 
i.e. the production of particles due to the presence of space-time 
varying external conditions has become subject of an increasing 
number of studies especially devoted to particle
creation in gravitational backgrounds (see for a review \cite{s4,bad82}), 
due to dynamical boundary conditions (see Refs. in \cite{spg97})
and photon production in moving dielectrics 
(see e.g. \cite{bae93,ebe96a,ebe96b}) 
to mention only a very few of prominent fields of research.

The phenomenon of quantum radiation
(QR) can be interpreted as the response of the vacuum of quantized fields
due to their interactions with external conditions. One may think of the
latter as some space-time dependent background fields (classical field
configuration) or as time-dependent, geometrical boundary conditions 
superimposed on the quantized fields under consideration 
(e.g. moving mirrors or dielectrics).

\bigskip
In a recent investigation \cite{spg97} we propose a canonical approach to the
dynamical Casimir effect in the presence of boundary conditions which
facilitates to calculate the number of generated particles by means of 
response theory. In this article we are going to apply this approach
to the situation of quantum fields in arbitrary space-time dependent background
fields lying particular emphasis on the situation of QED in moving and 
space-time varying dielectrics. Under rather general conditions the low-energy
behaviour of the spectrum of produced particles can be deduced. 
We also indicate that the formalism is most suitable for incorporating 
finite-temperature effects as well. This paper is organized as follows:
In section 2 we deduce some features of the energy spectrum of the dynamical
Casimir effect in smooth background fields based on the response theory approach.
In section 3 the formalism is applied to QED in dynamical dielectrics lying 
particular emphasis on the problem of field quantization. An interaction
Hamiltonian is derived which describes particle production due to two 
distinct generic prossess: 1. squeezing of the QED ground state due to localized,
space-time varying dielectric properties inside the medium; 2. velocity
effect due to motion of the medium. Section 4 examines the squeezing 
effect where a general expression for the total radiated 
energy will be derived.
Section 5 discusses the velocity effect qualitatively to underline
the importance of realistic ans\"atze for the velocity field and how they
may affect the energy-spectrum of the produced particles. We will close with
some discussions devoted to the following issues: generalization
of the formalism to finite temperatures, application to media with 
nontrivial permeability function and an indication of how to estimate  
the contribution of QR to the phenomenon of sonoluminescence. 

\section{Energy spectrum of quantum radiation}

In this section we present the general formalism for the calculation
of the number of particles produced by the dynamical Casimir effect.
We consider a set of quantized fields $\hat{\Theta}_A$ interacting with a 
classical background described by some smooth, space-time dependent
functions $\Delta K^{AB}_{\mu\nu}$. The rather general system under
consideration may be characterized by the four suppositions:

\begin{tabular}[t]{cll}
{\sl I  } &:& The $\hat{\Theta}_A$ are supposed to be massless and 
non-selfinteracting bosonic fields . 
\\
{\sl II } &:& The field configuration is considered at zero temperature \\
          & & and initially in the vacuum state.\\
{\sl III} &:& A proper definition of particles is provided by the
              unperturbed Hamiltonian $\hat{H}_0$ . \\
{\sl IV } &:& The interaction Hamiltonian $\hat{H}_1$ is determined by small, 
              space--time localized and \\
          & & smooth functions $\Delta K^{AB}_{\mu\nu}$.
\end{tabular}

These four assumptions allow to deduce the low-energy behaviour of the
spectrum $e(\omega)$ of the produced particles. 
We shall prove that for small freqencies $\omega$ the energy spectrum 
$e(\omega)$ will be proportional to $\omega^4$ in contrast to the
thermal spectrum which behaves as $\omega^3$. 

\bigskip

At first we specify a class of Lagrangians compatible with the four
assumptions stated above. 
Due to supposition {\sl I} we assume $\hat{\cal L}$ as a bilinearform 
of first order derivatives of the fields $\hat{\Theta}_A$ interacting
with background fields $\Delta K^{AB}_{\mu\nu}$:

\begin{equation}
\label{sumconv}
\hat{{\cal L}}=\frac{1}{2}\, \partial^\mu\hat{\Theta}_A
\left(K^{AB}_{\mu\nu}+\Delta K^{AB}_{\mu\nu}(\vec r,t)\right)
\partial^\nu\hat{\Theta}_B
\end{equation}

with $\Delta K^{AB}_{\mu\nu}(\vec r,t)\rightarrow 0$ 
for $|\vec r\,|\rightarrow\infty$ and for
$t\rightarrow\pm\infty$ (according to {\sl IV}).
In Eq. (\ref{sumconv}) and in the following formulae
we make use of the summation convention and declare that
one has to sum over all indices ($A$, $\mu$, $i$ etc.) that do
not occour at both sides of the equation. 

This leads to the total Hamiltonian :  

\begin{equation}
\hat H(t)=\hat H_0+\hat H_1(t)
=\int d^3r\left(\hat{\cal H}_0+\hat{\cal H}_1(t)\right)
\end{equation}

with $\hat{\cal H}_1(\vec r,t)\rightarrow 0$ 
for $|\vec r\,|\rightarrow\infty$ and for $t\rightarrow\pm\infty$ 
(according to {\sl IV}) together with the undisturbed Hamiltonian density

\begin{equation}
\label{Hnull}
\hat{\cal H}_0=\frac{1}{2}\hat\Pi_ A T^{AB} \hat\Pi_ B
+\frac{1}{2} V^{AB}_{ij}\partial^i\hat\Theta_ A\partial^j\hat\Theta_ B
\quad ,
\end{equation}

assuming the constants $K^{AB}_{\mu\nu}$ satisfy 
(possibly after a suitable transformation)
$K^{AB}_{0i}=K^{AB}_{i0}=0$ or other appropriate conditions
leading to the constant matrices $T^{AB}$ and $V^{AB}_{ij}$.

A proper definition of particles with
respect to the undisturbed Hamiltonian $\hat{H}_0$ (according to 
{\sl III}) is provided by insertion of the expansions

\begin{equation}
\label{feld}
\hat\Theta_ A(\vec r,t)=
V^{-1/2}F^B_A\sumint{\vec k}
\sqrt{\frac{1}{2\omega_{\vec k B}}}
(\hat a^+_{\vec k  B}(t)\;e^{i\vec k\vec r}
+{\rm h.c.})
\end{equation}

for the fields and

\begin{equation}
\label{moment}
\hat\Pi_ A(\vec r,t)=V^{-1/2}G^B_A\sumint{\vec k}
i\sqrt{\frac{\omega_{\vec k B}}{2}}
(\hat a^+_{\vec k  B}(t)\;e^{i\vec k\vec r}
-{\rm h.c.})
\end{equation}

for the canonical field momenta. 
($V$ denotes the quantization volume.) 

Supposition {\sl III} demands the existence of appropriate matrices
$F^A_B$ and $G^A_B$ which allow to diagonalize $\hat{H}_0$ 
(i.e. terms like $\hat a_{\vec k A}\hat a_{\vec k'B}$ 
and $\hat a^+_{\vec k A}\hat a^+_{\vec k'B}$ generated by the
insertion of (\ref{feld}) and (\ref{moment}) into (\ref{Hnull}) 
have to cancel) leading to the form

\begin{equation}
\hat H_0
=\sumint{\vec k  A}
\omega_{\vec k A}\left(\hat a^+_{\vec k  A}\hat a_{\vec k  A}+\frac{1}{2}\right)
=\sumint{\vec k  A}
\omega_{\vec k A}\left(\hat N_{\vec k  A}+\frac{1}{2}\right)\quad .
\end{equation}

The $\hat H_0$-frequencies $\omega_{\vec k A}$ possess a quasi-linear 
dispersion relation (after {\sl I}) :

\begin{equation}
\omega_{\vec k A}=\sqrt{k_i k_j W^{ij}_A}={\cal O}(k=|\vec k|)\quad .
\end{equation}

The following calculations are most suitably performed in 
interaction-representation:

\begin{equation} 
\frac{d\hat A}{dt}=
i[\hat H_0 , \hat A]+\frac{\partial\hat A}{\partial t} \quad ,
\end{equation}

\begin{equation} 
\label{psi}
\frac{d}{dt}\ket\psi=-i\hat H_1(t)\ket\psi  \quad .
\end{equation}

Suppositions {\sl II} and {\sl III} imply the following definition of 
the vacuum -- the initial state -- as well
as that of proper particle creation/annihilation operators 
$\hat a^+_{\vec k  A}$/$\hat a_{\vec k  A}$:

\begin{equation}
\ket{\psi(t\rightarrow-\infty)}=\ket{0}
\quad{\rm with}\quad
\forall\;\vec k A\quad\hat a_{\vec k  A}\ket{0}=0 \quad .
\end{equation}

Eq. (\ref{psi}) can be formally integrated with the 
time-ordering operator ${\cal T}$ :

\begin{eqnarray} 
\ket{\psi(t\rightarrow\infty )}
&=& {\cal{T}}
  \left[\exp\left(-i\int\,dt\,\hat H_1(t)\right)\right]\ket{0} \nonumber\\
&=& {\cal{T}}
  \left[\exp\left(-i\int\,d^4x\,\hat{\cal H}_1(\underline x)\right)\right]\ket{0} \nonumber\\
&=& \sum_{n=0}^\infty \,\frac{(-i)^n}{n!}\,
  \int dt_n \cdots \int dt_1\, {\cal T}\, 
  \left[\hat{H}_1(t_n)\cdots \hat{H}_1(t_1)\right] \ket{0} \nonumber\\ 
&=& \sum_{n=0}^\infty \,\frac{(-i)^n}{n!}\,
  \int d^4x_n \cdots \int d^4x_1\, {\cal T}\, 
  \left[\hat{\cal H}_1(\underline x_n)\cdots 
  \hat{\cal H}_1(\underline x_1)\right] \ket{0} 
\quad ,
\end{eqnarray}

with $\underline x=(\vec r,t)$. The expectation value of the
number opertator is given by

\begin{eqnarray}
\label{allgN}
\left<\hat N_{\vec k A}\right>
&=&
\bra{\psi(t\rightarrow\infty)}
\hat N_{\vec k A}
\ket{\psi(t\rightarrow\infty)} \nonumber\\
&=&
\sum_{m=1}^\infty \,\frac{(+i)^m}{m!}\,\int d^4x_m \cdots \int d^4x_1 
\sum_{n=1}^\infty \,\frac{(-i)^n}{n!}\,\int d^4x'_n \cdots \int d^4x'_1
\nonumber\\& & 
\bra{0}
{\cal T}\,\left[\hat{\cal H}_1(\underline x_m)\cdots 
\hat{\cal H}_1(\underline x_1)\right]
\hat N_{\vec k A}
{\cal T}\,\left[\hat{\cal H}_1(\underline x'_n)\cdots 
\hat{\cal H}_1(\underline x'_1)\right]
\ket{0}
\quad .
\end{eqnarray}

The vacuum expectation values have to be calculated in the
$\hat H_0$-dynamics and can also be deduced from the 
$2(n+m+1)$-point correlation functions
$\bra{0}\hat{\Theta}_{A_1}(\underline x_1)\cdots\hat{\Theta}_{A_p}(\underline x_p)\ket{0}$. 
Due to assumption {\sl I} the pertubation 
Hamiltonian density can be expressed as: 

\begin{equation}
\hat{\cal H}_1=Q^{AB}_{\mu\nu}(\underline x)
\partial^\mu\hat\Theta_ A\partial^\nu\hat\Theta_ B
\quad .
\end{equation}

In the interaction representation the time evolution 
of the fields is now governed by $\hat H_0$ and assumes the form:

\begin{equation}
\hat\Theta_ A(\underline x)=V^{-1/2}F^B_A\sumint{\vec k}
\sqrt{\frac{1}{2\omega_{\vec k B}}}
(\hat a^+_{\vec k B}\;e^{i\underline k_B\underline x}
+{\rm h.c.}) \quad ,
\end{equation}

with $\underline k_A=(\vec k,\omega_{\vec k A})$. 
The interaction term becomes

\begin{eqnarray}
\int d^4x\,\hat{\cal H}_1(\underline x)&=&
\sumint{\vec k_1\vec k_2}
\frac{k_{1C}^\mu k_{2D}^\nu}{2V\sqrt{\omega_{\vec k_1 C}\omega_{\vec k_2 D}}}
\times
\nonumber\\
& &
\int d^4x\,Q^{ A B}_{\mu\nu}(\underline x)F^C_A F^D_B
(\hat a^+_{\vec k_1  C}e^{i \underline k_1 \underline x}+{\rm h.c.})
(\hat a^+_{\vec k_2  D}e^{i \underline k_2 \underline x}+{\rm h.c.})\quad .
\end{eqnarray}

In order to obtain
a non-zero matrix-element in (\ref{allgN}) 
at the left and at the right of $\hat N_{\vec k A}$ 
at least one summation index $\vec k_{\ell}$ has to be equal to $\vec k$, 
which generates an overall factor of order $k$ . Assuming 
the functions $Q^{AB}_{\mu\nu}(\underline x)$ to be localized and smooth 
enough ({\sl IV}) , all mode summations are convergent and the 
$d^4x$-integrals are finite for the limit $k \rightarrow 0$.
Under these conditions, we can conclude that the number of particles
per mode possess the low-momentum behaviour

\begin{equation}
\left<\hat N_{\vec k A}\right>
={\cal O}(k) \quad .
\end{equation}

This enables us to calculate the total radiated energy associated with
the total number of created particles:

\begin{equation}
E=\sumint{\vec k A}\omega_{\vec k A}
\left<\hat N_{\vec k A}\right>
\rightarrow
\frac{V}{(2\pi)^3}
\sum\limits_{A}\int d^3k\;
\omega_{\vec k A}
\left<\hat N_{\vec k A}\right>
=\sum\limits_A \int\limits_0^\infty d \omega_A \, e_A(\omega_A)
=\int\limits_0^\infty d\omega\,e(\omega) \;\, .
\end{equation}

In the limiting case of small frequencies $\omega$
simple power counting leads to a non-thermal spectral energy density:

\begin{equation}
e(\omega)\sim\omega^4+{\cal O}(\omega^5)\quad .
\end{equation}

This reveals a general feature of the low-energy part of the spectrum of 
particles created by the dynamical Casimir effect due to the interaction
with external background fields, assuming that the suppositions 
{\sl I} -- {\sl IV} , which describe a very general case, 
are fulfilled.

\section{QED in dielectrics}

\subsection{Equations of motion}

Throughout this article natural units with 
$\hbar=c=\varepsilon_0=\mu_0=1$ will be used. We start with the
source free Maxwell equations ( i.e. for $\rho=0$ and $\vec j=\vec 0$ ) :

\begin{eqnarray}
\label{eq:maxw1}
\nabla\vec B&=0\quad ,\hspace{2cm}\nabla\times\vec E&={-\dot{\vec B}}
\quad ,\\
\nabla\vec D&=0\quad ,\hspace{2cm}\nabla\times\vec H&=\dot{\vec D}\quad .
\label{eq:maxw2}
\end{eqnarray}

The first pair (\ref{eq:maxw1}) leads to the definition of the potentials :

\begin{equation}
\vec B =-\nabla\times\vec A\quad ,\hspace{2cm}\vec E=\dotA+\nabla\Phi\quad .
\end{equation}

Now we suppose a linear and nondispersive medium in arbitrary motion 
described by the scalar dielectricity function $\varepsilon(\vec r,t)$ and 
a velocity field $\vec\beta(\vec r,t)$.
For the special case of a medium globally at rest, i.e. 
$u^\mu=(1,\vec \beta = \vec 0)$, where the relations 
$\vec D=\varepsilon\vec E$ and $\vec H=\vec B$ hold,
the Lagrangian leading to the second pair of the Maxwell equations reads:

\begin{equation}
\label{rest}
{\cal L}=\frac{1}{2}(\varepsilon\vec E^2-\vec B^2)\quad .
\end{equation}

There are different options for the construction of the Lagrangian for the 
general case of arbitary $u^\mu(\vec r,t)$ and $\varepsilon(\vec r,t)$.
The most direct way is to make an ansatz as a linear combination of all 
possible Lorentz-covariant scalars constructed from the tensors at hand:
the field strength tensor $F^{\mu\nu}$, the Lorentz metric $g_{\mu\nu}$,
the four velocity $u_\mu$, possibly the Levi-Civita pseudo-tensor
$\epsilon_{\mu\nu\rho\sigma}$ and the dielectricity function $\varepsilon$
(Lorentz-scalar). For reasons of simplicity one may neglect additional
acceleration effects that could be carried by $u_{\mu{\displaystyle,}\nu}$ 
or $\varepsilon_{{\displaystyle,}\mu}$ etc. 
The only form that reduces to the special case (\ref{rest}) of a medium at
rest and the vacuum Lagrangian 
${\cal L}=\frac{1}{4}F_{\mu\nu}F^{\nu\mu}$
for $\varepsilon=1$ , is provided by

\begin{eqnarray}
\label{allgLag}
{\cal L}
&=&\frac{1}{4}F_{\mu\nu}F^{\nu\mu}
+\frac{\varepsilon-1}{2}u^\mu F_{\mu\nu}F^{\nu\rho}u_\rho
\nonumber\\
&=&\frac{1}{2}(\vec E^2-\vec B^2)-\frac{\varepsilon-1}{2}
(1-\vec\beta^2)^{-1}
\left((\vec\beta\vec E)^2-(\vec E-\vec\beta\times\vec B)^2\right)
\quad .
\end{eqnarray}

This Lagrangian established by means of covariance arguments
may be expanded in powers in $\vec\beta$ (assuming $|\vec\beta|\ll 1$)
resulting in the non-covariant form

\begin{equation}
\label{lagr}
{\cal L}=\frac{1}{2}(\varepsilon\vec E^2-\vec B^2)
+(\varepsilon-1)\vec\beta(\vec E\times\vec B)
+{\cal O}(\vec\beta^2)            \quad .
\end{equation}

The Euler-Lagrange equations for the scalar potential $\Phi$ lead to:

\begin{equation}
\nabla\frac{\partial\cal L}{\partial\nabla\Phi}=
\nabla\vec D=0
\end{equation}

while the equations of motion for $\vec A$ read:

\begin{equation}
\frac{\partial}{\partial t}\frac{\partial\cal L}{\partial\dot{A^i}}=
\dot D_i=
-\frac{\partial}{\partial x^j}\frac{\partial\cal L}{\partial A^i_{,j}}=
(\nabla\times H)_i
\quad .
\end{equation}

So the Euler-Lagrange equations are identical with the second pair of the 
Maxwell equations, if we introduce the electric and magnetic displacement 
fields $\vec D$ and $\vec H$ according to

\begin{equation}
\vec D=\frac{\partial{\cal L}}{\partial\vec E}=
\varepsilon\vec E+(\varepsilon-1)\vec B\times\vec\beta
+{\cal O}(\vec\beta^2)
\label{dfield}
\end{equation}

and

\begin{equation}
\vec H=-\frac{\partial{\cal L}}{\partial\vec B}=
\vec B+(\varepsilon-1)\vec E\times\vec\beta 
+{\cal O}(\vec\beta^2)\quad .
\label{hfield}
\end{equation}

Applying Euler's theorem the Lagrangian (\ref{allgLag}) 
transforms into the alternative form

\begin{equation}
{\cal L}=\frac{1}{2}(\vec E\vec D - \vec B\vec H)\quad ,
\end{equation}

which holds to all orders in $\vec \beta$ .

\subsection{Gauge problems}

If we try to carry out the canonical quantization based on the 
Hamiltonian formalism applied to the Lagrangian obtained above, 
we are faced with some problems arising from the fact, that QED is a gauge
field theory. 
(Attempts for quantizing QED in dielectric media date back for more than
half of a century \cite{bai34,jaw48}. To solve the quantization problem 
for general media has attracted recent interest in quantum optics,
see \cite{kaw88}-\cite{bkv97}.)
Due to the fact, that $\cal L$ contains 
no kinetic term for the scalar field component $\Phi$ there exists the 
primary constraint:

\begin{equation}
\Pi_0=\frac{\partial\cal L}{\partial\dot\Phi}=0
\quad .
\end{equation}

A secondary constraint arises from the absence of
any potential term for the longitudinal component
of the vector potential $\vec A_\|$:

\begin{equation}
\nabla\vec\Pi=\nabla\frac{\partial\cal L}{\partial\dotA}=
\nabla\frac{\partial\cal L}{\partial\vec E}=
\nabla\vec D=0       \quad .
\end{equation}

So far these problems also occur in vacuum QED. However, in that case
it is possible to remove both kinds of field modes simultaneously
by choosing the Coulomb gauge $\nabla\vec A=0$ which has $\Phi =0$ as 
immediate consequence. Equivalently, one can choose the temporal
gauge $\Phi=0$ which directly leads to $\nabla\vec A=0$. 
For a detailed discussion of the problem of the quantization
with constraints we refer e.g. to \cite{gauge}.
So in vacuum QED it is possible to remove the scalar and the
longitudinal photons simultaneously. This fails however for QED in media 
because of the non-trivial relation between $\vec E$ and $\vec D$. 
As the next step we observe that the Hamiltonian derived according to the
canonical prescription

\begin{equation}
{\cal H}(\vec\Pi,\vec A, \Phi)=\vec\Pi\dotA-{\cal L}=
\frac{1}{2}(\vec E\vec D+\vec B\vec H)
-\vec\Pi\,\nabla\Phi
\label{naiv}
\end{equation}

contains the term $\vec\Pi\,\nabla\Phi$. Even though $\nabla\vec\Pi=0$, 
dropping this term or setting $\Phi =0$ 
would change the equations of motion, 
e.g. such that they are no more consistent with the Coulomb gauge 
$\nabla\vec A=0$. One way to solve the gauge problem is provided 
by carrying out the procedure of Dirac quantization as described e.g. 
in \cite{gauge}. One could also try to generalize the path integral quantization 
recently applied to dielectrics at rest \cite{bkv97}.
Here we proceed in a different way: Based on a decomposition of the electromagnetic
fields into an appropriatly choosen basis $\{\vec f_a\}$
we introduce a set of independent
variables $\{q_a\}$ leading to a Lagrangian $L(q,\dot q,t)$ without
any constraints and perform the usual canonical quantization.    
It will be necessary to prove that the primary and secondary constraints
required for the fields can be satisfied by the adequate choice for 
$\{\vec f_a\}$. 

For this purpose we fix the gauge by choosing the Coulomb condition:

\begin{equation}
\nabla\vec A=0 \quad .
\end{equation}

As mentioned above, in dielectrics this does not automatically 
imply $\Phi=0$. Furthermore, from the equation of motion
$\nabla\vec D=0$ it follows:

\begin{equation}
\label{phi}
\nabla(\varepsilon\nabla\Phi)
=-\nabla\left(
\varepsilon\dotA
+(\varepsilon-1)\vec\beta\times\left(\nabla\times\vec A\right)
\right)+{\cal O}(\vec\beta^2) \quad ,
\end{equation}

which possesses non-trivial solutions $\Phi \neq 0$.
Requiring the boundary conditions $\Phi\rightarrow 0$ for
$|\vec r\,|\rightarrow\infty$ 
Eq. (\ref{phi}) is solvable with the aid of the Green function
$G(\vec r,\vec r\,',t)$  
corresponding to the Laplace-Beltrami-type operator 
$(\nabla\varepsilon\nabla)$ :

\begin{eqnarray}
\label{philsg}
\Phi(\vec r,t)
&=&-\int d^3r'\;G(\vec r,\vec r\,',t)\times
\nonumber\\
& &
\nabla'\left(
\varepsilon(\vec r\,',t)\dotA(\vec r\,',t)+(\varepsilon(\vec r\,',t)-1)
\vec\beta(\vec r\,',t)\times\left(\nabla'\times\vec A(\vec r\,',t)\right)
\right)
+{\cal O}(\vec\beta^2)
\; .
\end{eqnarray}

This equation explicitly reveals the nature of the scalar potential
$\Phi$ as a dependent field variable. 
Since Eq. (\ref{phi}) represents an equation of motion 
the action $\cal A$ becomes extremal at the hypersurface described by  
(\ref{philsg}). Therefore the correct equations of motion could also be
obtained by varying $\cal A$ only on this hypersurface.
Accordingly, it is allowed to eliminate $\Phi$ in favour of the 
vector potential $\vec A$ by inserting Eq. (\ref{philsg})
into $L[\dotA,\vec A,\Phi,t]$ arriving at
$L'[\dotA,\vec A,t]=L\left[\dotA,\vec A,\Phi[\dotA,\vec A,t],t\right]$.

The next important step is to expand the remaining fields into an
appropriate basis set exploiting the gauge condition.
The irreducible degrees of freedom of $\vec A$ with $\nabla\vec A=0$
were described by an expansion into a complete set of real, orthonormal
and transverse functions $\vec f_a(\vec r\,)$ (see appendix):

\begin{equation}
\vec A(\vec r,t)=q_a(t) \vec f_a(\vec r\,) 
\quad , \quad
q_a(t)=\int d^3r\,\vec A(\vec r,t) \vec f_a(\vec r\,)\quad .
\end{equation}

Insertion of this expansion into (\ref{philsg}) allows to express
$\Phi$ completely in terms of the variables $q$ and $\dot q$ : 

\begin{eqnarray}
\Phi (\vec r,t) &=& q_a(t)\, \varphi_a(\vec r,t) + 
\dot q_a(t)\, \chi_a(\vec r,t) 
\quad ,
\\
\varphi_a(\vec r,t) &=& -\int d^3r'\;G(\vec r,\vec r\,',t) \nabla'\left(
(\varepsilon(\vec r\,',t)-1)
\stackrel{}{\vec\beta}(\vec r\,',t)\times
\left(\nabla'\times\vec f_a(\vec r\,')\right)\right)
+{\cal O}(\vec\beta^2)
\, ,
\\
\chi_a(\vec r,t) &=& -\int d^3r'\;G(\vec r,\vec r\,',t)
\nabla'\left(\varepsilon(\vec r\,',t)\vec f_a(\vec r\,')
\right)
+{\cal O}(\vec\beta^2)
\quad .
\end{eqnarray}

The scalar potential does no longer appear as a degree of freedom. 
We can express the physical fields:  

\begin{eqnarray}
\vec E(\vec r,t)&=&\dot q_a(t) \vec f_a(\vec r\,) + 
q_a(t) \nabla \varphi_a(\vec r,t) + \dot q_a(t) \nabla \chi_a(\vec r,t)
\quad , \label{efield} \\
\vec B(\vec r,t)&=&-q_a(t)\;\nabla\times\vec f_a(\vec r\,)\quad .
\label{bfield}
\end{eqnarray}

Insertion of these decompositions into the Lagrangian 
Eq. (\ref{allgLag}) leads to

\begin{eqnarray}
L(q,\dot q,t)
&=&
\int d^3r\;{\cal L}
=\int d^3r\left(\stackrel{}{\cal L}_0+{\cal L}_1(t)\right)
\nonumber\\
&=&
\frac{1}{2}(\varepsilon_\infty\dot q_a^2-\Omega_a^2q_a^2)
+q_a{\cal X}_{ab}(t)q_b
+q_a{\cal Y}_{ab}(t)\dot q_b
+\dot q_a{\cal Z}_{ab}(t)\dot q_b \quad .
\end{eqnarray}

Now we have a Lagrangian expressed by an independent set 
of variables $q$, that contains no constraints. 
The Euler-Lagrange  equations 

\begin{equation}
\frac{d}{dt}\frac{\partial L}{\partial\dot q_a}=
\frac{\partial L}{\partial q_a}
\end{equation}

lead to the Maxwell equation $\dot{\vec D}=\nabla\times\vec H$
after multiplication with $\vec f_a(\vec r\,)$ and reidentification 
of the fields. The other equations
-- especially $\nabla\vec D=0$ --  are automatically satisfied. 
Now we can derive the canonical momenta $p$ and their field 
representations $\vec\Pi$ (see appendix):

\begin{equation}
\vec\Pi=p_a\vec f_a=\frac{\partial L}{\partial\dot q_a}\vec f_a=
\left(\varepsilon\vec E+(\varepsilon-1)\vec B\times\vec\beta
+{\cal O}(\vec\beta^2) \right)_\perp 
=\vec D
\quad .
\end{equation}

It turns out that the canonical momentum is equal to $\vec D$  
also in this general case. 
This is not obvious because the terms including $\Phi$ 
contain also $\dot q$ ; but performing the calculations we see that  
the contribution from $\Phi(q,\dot q,t)$ to $p$ vanishes 
due to $\nabla\vec D=0$ (see appendix):

\begin{equation}
\frac{\delta}{\delta\dotA_\perp}
L\left[\dotA,\vec A,\Phi[\dotA,\vec A,],t\right]
\equiv
\vec f_a\frac{\partial}{\partial\dot q_a}L(q,\dot q,t)
=
\frac{\delta}{\delta\vec E}
L\left[\vec E,\vec B,t\right]
\quad .
\end{equation}

The Hamiltonian now becomes:

\begin{equation}
H(p,q,t)=p_a\dot q_a-L=\int d^3r\;{\cal H}(p,q,t)
\end{equation}

with

\begin{equation}
{\cal H}(p,q,t)
=\frac{1}{2}\left(\frac{1}{\varepsilon(\vec r,t)}\vec\Pi^2
+(\nabla\times\vec A)^2\right)
+\frac{\varepsilon(\vec r,t)-1}{\varepsilon(\vec r,t)}
\vec\beta(\vec r,t)\left(\vec\Pi\times(\nabla\times\vec A)\right)
+{\cal O}(\vec\beta^2)
\end{equation}

respectively

\begin{equation}
{\cal H}(p,q,t)=
\frac{1}{2}(\vec E\vec D+\vec B\vec H)
\quad .
\end{equation}

We have thus derived the Hamiltonian which indeed has the 
same form as the one obtained in the naive way (\ref{naiv}), 
except for the term
$\vec\Pi\,\nabla\Phi$. Now $\Phi$ and $\vec A_\|$ are 
no longer degrees of freedom and the gauge condition is
automatically satisfied. As a consequence the term 
$\vec\Pi\,\nabla\Phi$ disappears.

\subsection{Quantization}

Based on the Lagrangian $L(\dot q,q,t)$ obtained above it is
possible to carry out the procedure of the canonical
quantization via defining the following commutation relations:

\begin{eqnarray}
[\hat q_a(t),\hat q_b(t)]=[\hat p_a(t),\hat p_b(t)]=0 
\quad , \quad
[\hat q_a(t),\hat p_b(t)]=i\,\delta_{ab}
\quad .
\end{eqnarray}

Multiplication with the eigenfunctions $\{ f_a \}$ leads to the commutation 
relations for the fields:

\begin{eqnarray}
[\hat A^i(\vec r,t),\hat A^j(\vec r\,',t)]=
[\hat\Pi^i(\vec r,t),\hat\Pi^j(\vec r\,',t)]=0 
\end{eqnarray}

and

\begin{eqnarray}
\label{BundD}
[\hat A^i(\vec r,t),\hat\Pi^j(\vec r\,',t)]
=i\,\delta^{ij}_\perp(\vec r-\vec r\,')
\quad .
\end{eqnarray}

Eq. (\ref{BundD}) can be used to obtain a medium-independent
(i.e. for arbitary $\varepsilon(\vec r,t)$ and $\vec\beta(\vec r,t)$)
commutation relation between the fields 
$\vec B=-\nabla\times\vec A$ and $\vec D=\vec\Pi$ (see e.g. \cite{bai34}) :

\begin{equation}
[\hat B^i(\vec r,t),\hat D^j(\vec r\,',t)]
=i\,\epsilon^{ijk}\,\partial_k\,\delta(\vec r-\vec r\,')
\quad .
\end{equation}

In the quantization prescription based on the decomposition presented 
above the Coulomb gauge condition
(and also $\nabla\vec D=0$) appears as an operator 
identity. This leads to

\begin{equation}
\forall\;\ket{\psi},n\;:\quad\bra{\psi}(\nabla\hat{\vec A\;})^n\ket{\psi}=0
\quad ,
\end{equation}

i.e., the fixing of the gauge removes all longitudinal photons.
But for the scalar photons this is not the case for any variation
of $\varepsilon(\vec r,t)$ or $\vec\beta(\vec r,t)$ :
 
\begin{equation}
\bra{0}\hat\Phi^2(\vec r,t)\ket{0}\neq 0
\quad .
\end{equation}

\subsection{Interaction Hamiltonian}

Now we are in the position to apply the formalism presented 
in section 2.
After separating the undisturbed $\hat H_0$ , the interaction Hamiltonian  
can be cast into the form:

\begin{equation}
\hat{\cal H}_1
=\frac{1}{2}
\left(\frac{1}{\varepsilon(\vec r,t)}-\frac{1}{\varepsilon_\infty}\right)
\hat{\vec\Pi}^2
+\frac{\varepsilon(\vec r,t)-1}{\varepsilon(\vec r,t)}
\vec\beta(\vec r,t)
\left(\hat{\vec\Pi}\times(\nabla\times\hat{\vec A\,})\right)
+{\cal O}(\vec\beta^2) \quad.
\end{equation}

There are two terms giving rise to disturbances of the vacuum and thus
contributing to the production of photons: the first
term due to the change of the dielectric function $\varepsilon(\vec r,t)$
-- the squeezing term -- and the second one related to $\vec\beta(\vec r,t)$
-- the velocity term. 

The assumption of localized perturbations as stated in section 2 
requires  
$\varepsilon(\vec r,t)\rightarrow\varepsilon_\infty$ and 
$\vec\beta(\vec r,t)\rightarrow 0$ 
for $|\vec r\,|\rightarrow\infty$ and for $t\rightarrow\pm\infty$.

\section{Squeezing effect}

As indicated in section 2, the general form for the particle
production due to the disturbance $\hat{\cal H}_1$ turns out to be:

\begin{eqnarray}
\left<\hat N_{\vec k\lambda}\right>
&=&
\bra{\psi(t\rightarrow\infty)}
\hat N_{\vec k\lambda}
\ket{\psi(t\rightarrow\infty)}
\nonumber\\
&=&
\int d^4x \int d^4x'  
\bra{0} 
\hat{\cal H}_1(\underline x)
\hat N_{\vec k\lambda} 
\hat{\cal H}_1(\underline x')
\ket{0}
+
{\cal O}(\hat H_1^3)
=
N_{\vec k\lambda}
+
{\cal O}(\hat H_1^3)
\quad .
\end{eqnarray}

Now we shall evaluate  this expression for $\vec\beta=\vec 0$,
neglecting the effects of any matter flow,
i.e., for the pure squeezing contribution:

\begin{equation}
\hat{\cal H}_1(\vec r,t)
=\frac{1}{2}
\left(\frac{1}{\varepsilon(\vec r,t)}-\frac{1}{\varepsilon_\infty}\right)
{\hat{\vec\Pi}}^2(\vec r,t)
=\xi(\vec r,t)\;{\hat{\vec\Pi}}^2(\vec r,t)
\quad .
\end{equation}

The mode expansion in the $\hat H_0$-dynamics leads to:

\begin{equation}
\label{moden}
\hat{\vec\Pi}
=\frac{i}{\sqrt{V}}\sumint{\vec k\,\lambda}
\sqrt{\frac{\omega_{\vec k}}{2}}
\left(\hat a^+_{\vec k\,\lambda}\vec e_{\vec k\,\lambda}
\;e^{i\underline k\, \underline x}
\,-\,{\rm h.c.}\right)
\end{equation}

with $\underline k=(\vec k,\omega_{\vec k})=(\vec k,k/\sqrt{\varepsilon_\infty})$
and $\vec k\vec e_{\vec k\lambda}=0$ .
Inserting this expansion into $\hat H_1$ we arrive at :

\begin{eqnarray}
N_{\vec k\lambda}
=
\int d^4x \int d^4x'  
\bra{0} 
\!\!\!\!&\xi(\underline x)&\!\!\!\!\!\!\!
\sumint{\vec k_1\lambda_1}
\sumint{\vec k_2\lambda_2}
\frac{\sqrt{\omega_{\vec k_1}\omega_{\vec k_2}}}{2V}
\hat a_{\vec k_1\,\lambda_1}\vec e_{\vec k_1\,\lambda_1}
\hat a_{\vec k_2\,\lambda_2}\vec e_{\vec k_2\,\lambda_2}
e^{-i(\underline k_1+\underline k_2)\underline x}
\nonumber\\
\times
\;
\hat N_{\vec k\lambda} 
\!\!\!&\xi(\underline x')&\!\!\!\!\!\!\!
\sumint{\vec k_3\lambda_3}
\sumint{\vec k_4\lambda_4}
\frac{\sqrt{\omega_{\vec k_3}\omega_{\vec k_4}}}{2V}
\hat a^+_{\vec k_3\,\lambda_3}\vec e_{\vec k_3\,\lambda_3}
\hat a^+_{\vec k_4\,\lambda_4}\vec e_{\vec k_4\,\lambda_4}
e^{i(\underline k_3+\underline k_4)\underline x'}
\;
\ket{0}
\quad .
\end{eqnarray}

Performing all $\vec k\lambda$-summations we would obtain
a distribution-like response function that is related to the
$6$-point correlation function. Here we proceed in a different way 
and carry out the $d^4x$-integration which yields
a more compact form of the equation in terms of 
Fourier transformation ${\cal F}$:

\begin{equation}
{\cal F}\,:\,f(\underline x)\rightarrow
\widetilde f(\underline k)=\int d^4x\;f(\underline x)\;e^{i\underline k\,\underline x}
\quad .
\end{equation}

Accordingly, we obtain 

\begin{eqnarray}
N_{\vec k\lambda}
&=&  
\sumint{\vec k_1\lambda_1}
\sumint{\vec k_2\lambda_2}
\sumint{\vec k_3\lambda_3}
\sumint{\vec k_4\lambda_4}
\frac{\sqrt{\omega_{\vec k_1}\omega_{\vec k_2}
            \omega_{\vec k_3}\omega_{\vec k_4}}}{4V^2}
\widetilde\xi^*(\underline k_1+\underline k_2)
\widetilde\xi(\underline k_3+\underline k_4)
\times
\nonumber\\
& &
(\vec e_{\vec k_1\,\lambda_1}
\vec e_{\vec k_2\,\lambda_2})
(\vec e_{\vec k_3\,\lambda_3}
\vec e_{\vec k_4\,\lambda_4})
\bra{0}
\hat a_{\vec k_1\,\lambda_1}
\hat a_{\vec k_2\,\lambda_2}
\hat N_{\vec k\lambda} 
\hat a^+_{\vec k_3\,\lambda_3}
\hat a^+_{\vec k_4\,\lambda_4}
\ket{0}
\end{eqnarray}

and evaluate the expectation values:

\begin{eqnarray}
\label{bsp}
N_{\vec k\lambda}=  
\sumint{\vec k'\lambda'}
\frac{\omega_{\vec k}\omega_{\vec k'}}{V^2}
|\widetilde\xi(\underline k+\underline k')|^2
(\vec e_{\vec k\,\lambda}
\vec e_{\vec k'\,\lambda'})^2
\quad .
\end{eqnarray}

Requiring again $\widetilde\xi(\underline k)$ to be
finite for $\vec k\rightarrow\vec 0$ and integrable
(see section 2)
power counting reveals that the number of
particles per mode possess a leading term 
$N_{\vec k\lambda}\sim\omega_{\vec k}$ 
at low energies leading to a non-thermal spectrum
$e(\omega)={\cal O}(\omega^4)$ also in this case.

With the aid of the completeness relation for the 
polarisation vectors $\vec e_{\vec k \lambda}$ and 
$\vec e_{\vec k}=\vec k/k$ :

\begin{equation}
\vec e_{\vec k}\otimes\vec e_{\vec k}
+\sum\limits_\lambda
\vec e_{\vec k \lambda}\otimes\vec e_{\vec k \lambda}
={\bf 1}\hspace{-4pt}1
\end{equation}

we arrive at

\begin{eqnarray}
\label{minus}
N_{\vec k\lambda}=  
\sumint{\vec k'}
\frac{\omega_{\vec k}\omega_{\vec k'}}{V^2}
|\widetilde\xi(\underline k+\underline k')|^2
\left(1-(\vec e_{\vec k'}\vec e_{\vec k\lambda})^2\right)
\quad .
\end{eqnarray}

For reasons of simplicity we consider radialsymmetric functions 
$\xi(\vec r,t)=\xi(r,t)$ and expand the Fourier transform $\widetilde\xi$
into a Taylor-series:

\begin{equation}
\label{mellin}
\widetilde\xi(\underline k)
=\int dt\,e^{i\omega t}\int d^3r\;e^{i\vec k\vec r}\;\xi(r,t)
=\int dt\,e^{i\omega t}\sum\limits_{n=0}^\infty{\cal M}_n(t)\,(\vec k^2)^n
=\sum\limits_{n=0}^\infty\widetilde{\cal M}_n(\omega)\,(\vec k^2)^n
\end{equation}

assuming that the ${\cal M}_n(t)$ which are related to the Mellin
transform of $\xi(r,t)$ exist, i.e. that $\xi(r,t)$ is 
sufficiently localized.

The zero term is proportional to the volume $\cal V$ of the disturbance.
For example for a bubble described by
$\varepsilon(\vec r,t)=\varepsilon_\infty+
(1-\varepsilon_\infty)\Theta(R(t)-r)$
it follows: 
${\cal M}_0(t)=\frac{1}{2}(1-\varepsilon^{-1}_\infty){\cal V}(t)$ .

Inserting Eq. (\ref{mellin}) into Eq. (\ref{minus}) leads to:

\begin{equation}
N_{\vec k\lambda}=  
\sumint{\vec k'}
\frac{\omega_{\vec k}\omega_{\vec k'}}{V^2}
\left(1-(\vec e_{\vec k'}\vec e_{\vec k\lambda})^2\right)
\sum\limits_{n=0}^\infty
\sum\limits_{m=0}^\infty
\widetilde{\cal M}_n^*(\omega_{\vec k}+\omega_{\vec k}')
\,\widetilde{\cal M}_m(\omega_{\vec k}+\omega_{\vec k}')
\,(\vec k+\vec k')^{2(n+m)}
\; .
\end{equation}

For the radialsymmetric case $N_{\vec k\lambda}$ does not depend on 
$\lambda$ and $\vec e_{\vec k}$ , therefore
we turn to 
$N_k=\sum\limits_\lambda N_{\vec k\lambda}$:

\begin{equation}
N_k=  
\sumint{\vec k'}
\frac{\omega_{\vec k}\omega_{\vec k'}}{V^2}
\left(1+(\vec e_{\vec k}\vec e_{\vec k'})^2\right)
\sum\limits_{n=0}^\infty
\sum\limits_{m=0}^\infty
\widetilde{\cal M}_n^*(\omega_{\vec k}+\omega_{\vec k}')
\,\widetilde{\cal M}_m(\omega_{\vec k}+\omega_{\vec k}')
\,(\vec k+\vec k')^{2(n+m)}
\quad .
\end{equation}

Evaluating  the total radiated energy

\begin{equation}
E=\sumint{\vec k}\omega_{\vec k}\,N_k
\rightarrow
\frac{V}{(2\pi)^3}\int d^3k\,\frac{k}{\sqrt{\varepsilon_\infty}}N_k
\end{equation}
  
we arrive after some calculations at

\begin{equation}
E=\sum\limits_{n=0}^\infty\sum\limits_{m=0}^\infty
\int dt 
\left(\left(\frac{d}{dt}\right)^{4+2n}{\cal M}_n(t)\right)
{\cal G}^{nm}
\left(\left(\frac{d}{dt}\right)^{4+2m}{\cal M}_m(t)\right)
\end{equation}

with a geometry-independent matrix 
${\cal G}^{nm}={\cal G}^{nm}(n+m,\varepsilon_\infty)$
(see appendix).

The lowest (volume) term becomes for the bubble example

\begin{equation}
E^{00}={\cal G}^{00}\int dt\,\left({\cal M}^{(4)}_0(t)\right)^2=
\left(\frac{\varepsilon_\infty}{2\pi}\right)^3
\frac{1}{3\cdot5\cdot7}
\left(\frac{1-\varepsilon_\infty^{-1}}{2}\right)^2
\int dt\,\stackrel{....\,}{\cal V}^2(t)
\quad .
\end{equation}

If we assume that the dynamics of the disturbance can be 
described by a maximal characteristic length scale 
$R_{\scriptscriptstyle\rm  max}$ 
and minimal time scale $T_{\scriptscriptstyle\rm  min}$ ,
this term is of order 
${\cal O}(R^6_{\scriptscriptstyle\rm  max}T^{-7}_{\scriptscriptstyle\rm  min})$.
Further terms contain additional factors of order 
${\cal O}(R^2_{\scriptscriptstyle\rm  max}T^{-2}_{\scriptscriptstyle\rm  min})$
and can be neglected for the case 
$R_{\rm\scriptscriptstyle  max}/T_{\scriptscriptstyle\rm  min}\ll 1$.

\section{Velocity effect}

As one can see in section 2, the spectrum of the photons
produced by the dynamical Casimir effect depends very strongly
on the concrete form of the disturbance.
Now we shall study some examples for the profile of the
velocity $\vec\beta(\vec r,t)$:\\
The first example corresponds to the radialsymmetric
flow of an incompressible fluid around a bubble with 
the time dependent radius $R(t)$:

\begin{equation}
\vec\beta(\vec r,t)=\dot R(t) \; \vec e_r \; R^2(t)/r^2
\quad .
\end{equation}

The Fourier transform $\widetilde{\vec\beta\;}(\vec k,\omega_{\vec k})$
remains finite in the limit $\vec k\rightarrow\vec 0$ and is of
order $R^3_{\scriptscriptstyle\rm  max}$. 
Therefore the spectrum of the created
photons behaves as $\omega^4+{\cal O}(\omega^5)$
for smooth and localized functions $R(t)$ and the velocity contribution is
for this case of the same order as the squeezing term.
(The perturbation Hamiltonian for the squeezing effect is also of order 
$R^3_{\scriptscriptstyle\rm  max}$ due to the $d^3r$-integration.) 
For the incompressible case the flow $\oint\vec\beta d\vec A$
of the fluid is instantaneously constant
even for $|\vec r\,|\rightarrow\infty$.
If we assume a more realistic situation, where $\vec\beta(\vec r,t)$ 
yields relevant contributions only over a bounded 
volume of order $R^3_{\scriptscriptstyle\rm  max}$ , 
the velocity effect turns out to be of order  
$R^4_{\scriptscriptstyle\rm  max}$ and can therefore be neglected in those
cases, where $R_{\scriptscriptstyle\rm  max}$ is sufficiently small. 
(The   ${\cal O}(R^4_{\scriptscriptstyle\rm  max})$-behaviour 
results from the $d^3r$-integration 
that implies ${\cal O}(R^3_{\scriptscriptstyle\rm  max})$ and the assumption, 
that $\vec\beta(\vec r,t)$ is still of order $R_{\scriptscriptstyle\rm  max}$ .)

For the second example:

\begin{equation}
\vec\beta(\vec r,t)=\dot R(t)\;\vec e_r
\end{equation}

that is studied in Ref. \cite{ebe96b}, one obtains a Fourier transform
that is divergent for  $\vec k\rightarrow\vec 0$.
This demonstrates the highly nonlocalized character of this profile
which results in an increasing flow for $|\vec r\,|\rightarrow\infty$.
Of course, for that case one could not provide the prediction 
$e(\omega)\sim\omega^4+{\cal O}(\omega^5)$.

For a third example we study the situation of a rigidly moving medium with

\begin{equation}
\varepsilon(\vec r,t)=\varepsilon_\infty
\quad{\rm and}\quad
\vec\beta(\vec r,t)=\vec\beta(t)
\quad .
\end{equation}
 
For that case Eq. (\ref{phi}) for $\Phi$ is easyly solvable:
$\Phi=(\varepsilon_\infty^{-1}-1)\vec\beta\vec A +{\cal O}(\vec\beta^2)$.
Calculating the perturbation Hamiltonian $\hat H_1$ we find no
photon production to first order in $\vec\beta$ (nonrelativistic limit):
$\hat H_1\ket{0}={\cal O}(\vec\beta^2)$ .
This implies that for the case of a purely time-varying velocity
the effect will be of order $N_{\vec k\lambda}={\cal O}(\vec\beta^4)$.

\section{Discussion}

\subsection{Finite temperatures}

The response theory approach to quantum radiation allows for
the incorporation of finite temperature effects. Instead of using the 
time evolution for the pure quantum state Eq. (\ref{psi}) one has to
consider the equation of motion for the statistical operator $\hat{\rho}$:

\begin{equation} 
\frac{d\hat\rho}{dt}=
-i[\hat H_1 , \hat\rho]\quad ,
\end{equation}

and Eq. (\ref{allgN}) becomes

\begin{eqnarray}
\left<\hat N_{\vec k A}\right>=
{\rm Tr}\left\{
\hat\rho_0
{\cal T}\left[\exp\left(i\int\,d^4x\,\hat{\cal H}_1
(\underline x)\right)\right]
\hat N_{\vec k A}
{\cal T}\left[\exp\left(-i\int\,d^4x'\,\hat{\cal H}_1
(\underline x')\right)\right]
\right\}\quad .
\end{eqnarray}

The zero-order term contains the undisturbed spectrum arising from 
$\hat\rho_0$.
The contribution of first order in $\hat H_1$ vanishes, if we assume a
$\hat\rho_0$ related to the thermodymamic equilibrium:

\begin{equation}
\hat\rho_0=\hat\rho(t\rightarrow-\infty)=
\exp(-\beta\hat H_0)/Z_0
\quad .
\end{equation}

The terms of second order contain the production of particles 
and the interaction between the thermal radiation field and the 
disturbance.

\subsection{Permeable media}

The approach presented above is generalizable to the case of media
described by $\varepsilon(\vec r,t)$, $\mu(\vec r,t)$ and
$\vec\beta(\vec r,t)$, if one adds a term proportional to 
$(\mu-1)\;u_\mu \stackrel{*}{F^\mu}_\nu\stackrel{*}{F^\nu}_\rho u^\rho$
with the dual tensor 
$\stackrel{*}{F}_{\mu\nu}=\epsilon_{\mu\nu\rho\sigma} F^{\rho\sigma}$
to the Lagrangian.

For pure magnetic disturbances ($\varepsilon=1$) the calculations 
proceed in the same way, if one replaces the potentials $\vec A$ and $\Phi$ 
by the dual potentials $\vec\Lambda$ and $\Gamma$ (see e.g. \cite{ham84}), defined via:

\begin{equation}
\vec D=\nabla\times\vec\Lambda
\quad , \quad
\vec H=\dot{\vec\Lambda}+\nabla\Gamma 
\quad .
\end{equation}

A possible application could be the case of magnetic phase transitions,
e.g. in connection to superconductivity.

\subsection{Contribution to sonoluminescence}

There have been various attempts to interprete the phenomenon of 
sonoluminescence \cite{bea97} as an effect of quantum radiation
\cite{sch,ebe96a,ebe96b}. In order to estimate the contribution 
of QR it is important to determine the characteristic parameters
(time- and lenght scales) involved in the dynamics of the bubble 
\cite{gea97,sah97}.
Calculations of the stationary Casimir energy 
(see Refs. \cite{cmpv96} and \cite{man97}) cannot account
for the dynamical character of the effect.
Without an upper limit for the volume $R^3_{\scriptscriptstyle\rm  max}$
and the time interval $T_{\scriptscriptstyle\rm  max}$ of the
disturbance and a frequency cutoff $K_c$ the radiated energy could 
diverge due to the space-time integration and the
mode summation, if the perturbation functions are not localized and smooth.
But if we assume a localized disturbance and a frequency cutoff due to the
fact, that $\varepsilon$ differs only from $1$ 
for wave numbers $k<K_c$, an estimation of the upper limit for the 
order of magnitude is possible.
We envisage Eq. (\ref{bsp}) as an example for the general case of the 
quadratic response, assuming that the higher order terms are much smaller:

\begin{eqnarray}
N_{\vec k\lambda}=  
\sumint{\vec k'\lambda'}
\frac{\omega_{\vec k}\omega_{\vec k'}}{V^2}
|\widetilde\xi(\underline k+\underline k')|^2
(\vec e_{\vec k\,\lambda}
\vec e_{\vec k'\,\lambda'})^2=  
\sumint{\vec k'\lambda'}
\frac{k'k}{V^2}
\int\!d^4x\!\int\!d^4x'
\;\Xi(\vec k,\lambda,\vec k',\lambda',\underline x,\underline x')
\end{eqnarray}

with a function $\Xi$ that is of order $1$.
The order of magnitude of the space-time integrations and the mode sum
can be estimated using $R_{\scriptscriptstyle\rm  max}$ , 
$T_{\scriptscriptstyle\rm  max}$ , $K_c$ and the quantization volume $V$:

\begin{equation}
\int d^4x\rightarrow 
{\cal O}(T_{\scriptscriptstyle\rm  max}R^3_{\scriptscriptstyle\rm  max}) 
\quad {\rm and} \quad
\sumint{\vec k}\rightarrow {\cal O}(VK^3_c)
\quad .
\end{equation}

This leads to $N^{\rm max}_{\vec k\lambda}={\cal O}
(T^2_{\scriptscriptstyle\rm  max}R^6_{\scriptscriptstyle\rm  max}K^5_c/V)$
which depends still on the quantization volume $V$.
But in the total energy as a physical observable $V$ of course does not enter: 

\begin{equation}
E^{\rm max}=\sumint{\vec k\lambda}
\omega_{\vec k}N_{\vec k\lambda}^{\rm max}
\rightarrow{\cal O}
(R^6_{\scriptscriptstyle\rm  max}T^2_{\scriptscriptstyle\rm  max}K^9_{c})
\quad .
\end{equation}

Another result of this approach is the fact that the Casimir contribution
to sonoluninescence should display a $\omega^4$-spectrum for small $\omega$
if we assume that the suppositions of section 2 are fulfilled.
This provides a possibility to distinguish this effect from others 
(e.g. photon emission due to heating the gas in the bubble) which
behave like  $\omega^3$.

\section*{Acknowledgement}

G. S. acknowledges support by BMBF, DFG and GSI.

\section*{Appendix}

We have to introduce a complete set of real, orthonormal
and transverse eigenfunctions $\vec f_a(\vec r\,)$ of the Laplace operator: 

\begin{equation}
\nabla^2 \vec f_a(\vec r\,)=-\Omega_a^2 \vec f_a(\vec r\,)\quad ,
\end{equation}

\begin{equation}
\forall\,a\quad\nabla\vec f_a(\vec r\,)=0\quad ,
\end{equation}

\begin{equation}
\int d^3r \vec f_a(\vec r\,) \vec f_b(\vec r\,)=\delta(a,b)\quad ,
\end{equation}

\begin{equation}
\sumint{a} f^i_a(\vec r\,) f^j_a(\vec r\,')=
\delta^{ij}_\perp(\vec r-\vec r\,')\quad .
\end{equation}

Accordingly, the field modes also satisfy the relation:

\begin{equation}
\int d^3r\;\left(\nabla\times\vec f_a(\vec r\,)\right)
           \left(\nabla\times\vec f_b(\vec r\,)\right)
         =\Omega_a^2\delta_{ab}\quad .
\end{equation}

Let us show how to calculate the canonical momenta $p$ as well as 
their field representation $\vec\Pi=p_a\vec f_a$ from the Lagrangian:

\begin{eqnarray}
p_a
&=&
\frac{\partial L}{\partial\dot q_a}=
\int d^3r\;\frac{\partial{\cal L}}{\partial\dot q_a}=
\int d^3r\;\frac{\partial{\cal L}}{\partial\vec E}
         \frac{\partial\vec E}{\partial\dot q_a}
\nonumber\\
&=&
\int d^3r\;\vec D(\vec r\,)
\left(\vec f_a(\vec r\,) + \nabla\chi_a(\vec r,t)\right)=
\int d^3r\;\vec D(\vec r\,)\vec f_a(\vec r\,)
\quad .
\end{eqnarray}

For the same reason $\Phi$ yields no direct contribution to
$\frac{\partial L}{\partial q_a}$ , i.e., to $\nabla\times\vec H$.

\begin{eqnarray}
\vec\Pi(\vec r\,)=
p_a\vec f_a(\vec r\,)
&=&
\int d^3r'\;\vec f_a(\vec r\,)\left(\vec f_a(\vec r\,')\vec D(\vec r\,')\right)
\nonumber\\
&=&
\int d^3r'\;\vec e_i\;\delta^{ij}_\perp(\vec r-\vec r\,')D_j(\vec r\,')=
\vec D_\perp(\vec r\,)=\vec D(\vec r\,)
\end{eqnarray}

The transverse part $\vec F_\perp(\vec r,t)$ of an arbitrary vector field 
$\vec F(\vec r,t)=\vec F_\|(\vec r,t)+\vec F_\perp(\vec r,t)$ is defined as

\begin{equation}
\nabla^{2}\vec F_\perp(\vec r\,)=
-\nabla\times\left(\nabla\times\vec F(\vec r\,)\right)
\quad .
\end{equation}

As a complete and orthonormal set of real and transversal modes we use

\begin{equation}
f_{\vec k \lambda + }(\vec r\,)=
\sqrt{\frac{2}{V}}\cos(\vec k\vec r\,)\vec e_{\vec k\lambda}
\end{equation}

and 

\begin{equation}
f_{\vec k \lambda - }(\vec r\,)=
\sqrt{\frac{2}{V}}\sin(\vec k\vec r\,)\vec e_{\vec k\lambda}
\end{equation}

with polarization vectors
$\vec e_{\vec k\lambda}=\vec e_{-\vec k\lambda}$. The $q$ 
are related to the creation/annihilation operators 
used in Eq. (\ref{moden}), e.g. for $q_{\vec k\lambda+}$ it follows:

\begin{equation}
\hat q_{\vec k \lambda + }=
\frac{1}{2\sqrt{\omega_{\vec k}}}
(
\hat a^+_{\vec k\lambda}
+\hat a_{\vec k\lambda}
+\hat a^+_{-\vec k\lambda}
+\hat a_{-\vec k\lambda}
)
\quad .
\end{equation}

To be complete we note here the explicit expression for the
Matrix ${\cal G}^{nm}$ used in section 4:

\begin{eqnarray}
{\cal G}^{nm}=\frac{\varepsilon_\infty^{3+n+m}}{(2\pi)^3(8+2n+2m)!}
\sum\limits_{\ell=0}^{n+m}
\left(\begin{array}{c}
{m+n}\\
{\ell}
\end{array}\right)
2^\ell\left((-1)^\ell+1\right)\left(\frac{1}{\ell+1}+\frac{1}{\ell+3}\right)
\times
\nonumber\\
\sum\limits_{s=0}^{n+m-\ell}
\left(\begin{array}{c}
{m+n-\ell}\\
{s}
\end{array}\right)
(4+\ell+2s)!(3-\ell-2s+2n+2m)!
\quad .
\end{eqnarray}

\end{document}